\documentclass[prl,aps,amssymb,nofootinbib,twocolumn,showpacs]{revtex4}
\usepackage{amsmath}
\usepackage{amssymb}
\usepackage{amsthm}
\usepackage{amsfonts}
\usepackage{enumerate}
\usepackage{latexsym}
\usepackage[dvips]{graphicx}

\newcommand{\beq}{\begin{equation}}
\newcommand{\eneq}{\end{equation}}
\newcommand{\beqs}{\begin{equation*}}
\newcommand{\eneqs}{\end{equation*}}

\input{epsf}

\begin{document}

\tolerance 1000

\title{Elementary Excitations of Quantum Critical 2+1 D
Antiferromagnets}

\author { Zaira Nazario$^\dagger$ and 
David I. Santiago$^{\dagger, \star}$ }

\affiliation{$\dagger$ Department of Physics, Stanford University,
             Stanford, California 94305 \\ 
	     $\star$ Gravity Probe B Relativity Mission, Stanford,
	     California 94305}
\begin{abstract}
\begin{center}

\parbox{14cm}{ It has been proposed that there are degrees of freedom
intrinsic to quantum critical points that can contribute to quantum
critical physics. We point out that this conclusion is quite general
below the upper critical dimension. We show that in $2+1$ D
antiferromagnets skyrmion excitations are stable at criticality and
identify them as the critical excitations. We found exact solutions
composed of skyrmion and antiskyrmion superpositions, which we call
topolons. We include the topolons in the partition function and
renormalize by integrating out small size topolons and short
wavelength spin waves. We obtain correlation length exponent
$\nu=0.9297$ and anomalous dimension $\eta=0.3381$.}

\end{center}
\end{abstract}
\date{\today}

\maketitle

Quantum phase transitions have been a subject of theoretical and
experimental exploration since the pioneering work of John
Hertz\cite{hertz}. Since then, quantum critical behavior has been
understood and studied as arising from quantum fluctuations of the
order parameter\cite{hertz,millis}. In this traditional approach the
quantum transition is studied via the Wilson renormalization group in
which fluctuations of the order parameter are taken properly into
account. It is said that quantum phase transitions follow the
Landau-Ginzburg-Wilson paradigm (LGW).

It has recently been suggested that quantum critical points will have
properties that cannot be obtained from LGW order parameter
fluctuations alone\cite{bob1,sachdev2}. In particular, it was
suggested that quantum critical points will have low energy elementary
excitations intrinsic to the critical point whose fluctuations will
contribute and modify the critical properties. It was postulated that
these excitations will be fractionalized\cite{bob1,sachdev2}. 

That some quantum critical points have elementary excitations
different from those of each of the phases it separates can be
inferred quite generally. We concentrate in relativistic quantum
critical points, but we emphasize that these physics can take place in
other systems. For such a system, which we take to be an
antiferromagnet, we are interested in the N\`eel magnetization Green's
function, or staggered magnetic susceptibility.

In the ordered phase the transverse Green function or susceptibility
corresponds to spin wave propagation and it has a nonanalyticity in
the form of a pole corresponding to such propagation:
    \beq
      \nonumber
      \langle \vec n(-\omega, -\vec k) \cdot \vec n(\omega, \vec
      k)\rangle = \frac{Z(\omega, \vec k)}{c^2k^2 - \omega^2} +
      G_{\text{incoh}} (\omega, \vec k) \,.
    \eneq
Here $Z(\omega, \vec k)$ is between 0 and 1, and the incoherent
background $G_{\text{incoh}}$ vanishes at long wavelengths and small
frequencies. The fact that the Green's function has a pole means that
transverse Goldstone spin waves are low energy eigenstates of the
antiferromagnet. At criticality, the system has no N\`eel order and
thus Goldstones cannot be elementary excitations of the system.

In the disordered phase the Green function or susceptibility
corresponds to spin wave propagation with all three polarizations and
it has a pole nonanalyticity corresponding to such propagation:
    \beq 
      \nonumber
      \langle \vec n(-\omega, -\vec k) \cdot \vec n(\omega, \vec
      k)\rangle = \frac{A(\omega, \vec k)}{c^2k^2 + \Delta^2-
      \omega^2} + G_{\text{incoh}} (\omega, \vec k) \,.
    \eneq
Here $A(\omega, \vec k)$ is between 0 and 1, and the incoherent
background $G_{\text{incoh}}$ vanishes at long wavelengths and small
frequencies, $\Delta$ is the gap to excitations in the disordered
phase. That this Green's function has a pole means that triplet or
triplon spin waves are low energy eigenstates of the disordered
antiferromagnet. For 2+1 D antiferromagnets, and in general for
antiferromagnets below the upper critical dimension, the quasiparticle
pole residue $A$ vanishes as the system is tuned to the quantum
critical point\cite{halp}. At criticality, triplon excitations
have no spectral weight and thus triplons cannot be elementary
excitations of the system.

On the other hand right at criticality the response function below the
upper critical dimension (below which $\eta \neq 0$, while above
$\eta=0$) has nonanalyticities that are worse than poles
    \beq 
    \label{qresp}
      \langle \vec n(-\omega, -\vec k) \cdot \vec n(\omega, \vec
      k)\rangle = A'\left(\frac{1}{c^2k^2 - \omega^2} \right)^{1 -
      \eta/2}
    \eneq
as obtained from the renormalization group studies of the nonlinear
sigma model\cite{halp,polya2}. Below the upper critical dimension
$\eta$ is a nonintegral universal number for each dimensionality. This
critical susceptibility has no pole structure, but has a branch
cut. It sharply diverges at $\omega=ck$ and is purely imaginary for
$\omega>ck$. Branch cuts in quantum many-body or field theory
represent immediate decay of the quantity whose Green function is
being evaluated. Hence the elementary excitations or eigenstates of
the noncritical quantum mechanical phases break up as soon as they are
produced when the system is tuned to criticality: they do not have
integrity. The complete lack of pole structure and the branch cut
singularity below the upper critical dimension mean that the
elementary excitations of the quantum mechanical phases away from
criticality, the spin waves, {\it cannot even be approximate
eigenstates} at criticality as they are absolutely unstable.

The quantum critical point is a unique quantum mechanical phase of
matter, which under any small perturbation becomes one of the phases
it separates. It is a repulsive fixed point of the renormalization
group. As far as the transition from one quantum mechanical phase to
the other is continuous, and both phases have different physical
properties, the critical point will have its unique physical
properties different from the phases it separates. The properties of
the critical point follow from the critical Hamiltonian $H(g_c)$
($g_c$ is the critical coupling constant), which will have a unique
ground state and a collection of low energy eigenstates which are its
elementary excitations. These low energy eigenstates are different
from those of each of the phases as long as we are below the upper
critical dimension. {\it As a matter of principle, all quantum
critical points below the upper critical dimension will have their
intrinsic elementary excitations}.

We have seen that below the upper critical dimension, the excitations
of the stable quantum phases of the system become absolutely unstable
and decay when the system is tuned to criticality. The question comes
to mind immediately: what could they be decaying into? When one tries
to create an elementary excitation of one of the phases, it will decay
immediately into the elementary excitations of the critical point. The
critical excitations will be bound states of the excitations of the
stable phases the critical point separates. These bound states could
be fractionalized as conjectured by Laughlin\cite{bob1} and Senthil,
{\it et. al.}\cite{sachdev2}, but they need not be in all cases. These
critical degrees of freedom are responsible for corrections to the LGW
phase transition canon\cite{sachdev2}. {\it The intrinsic quantum
critical excitations contribute to the thermodynamical and/or physical
properties of the quantum critical system.}

Now we turn to a specific model in order to identify what the critical
excitations are. We concentrate on $2+1$ dimensional short-range
Heisenberg antiferromagnets in a bipartite lattice. These are
described by the $O(3)$ nonlinear sigma model augmented by Berry
phases\cite{hal2,subir}
    \begin{align}
      \begin{aligned}
	\label{action}
	\mathcal Z &=\int \mathcal D\vec n \delta(\vec n^2-1)
	e^{-\mathcal S_E} \\ 
	\mathcal S_E &= iS_B + \int_{0}^{\beta} \! d\tau \!  \int \!
	d^2\vec x \frac{\rho_s}{2} \left[(\partial_{\vec x}\vec n)^2 +
	\frac{(\partial_\tau\vec n)^2}{c^2} \right] \,.
      \end{aligned}
    \end{align}
where $\rho_s\equiv J S^2$ is the spin stiffness, and the spin-wave
velocity $c=2 \sqrt{2} JSa$ with $a$ the lattice constant and $J$ the
microscopic spin exchange. The Berry phase terms represent the sums of
the areas swept by the vectors $\vec n_i(\tau)$ on the surface of a
unit sphere as they evolve in Euclidean time\cite{hal2}. They were
shown to be zero\cite{hal2,nohopf} in the N\`eel phase and the
critical point, but relevant in the disordered phase\cite{sachread}.
We drop the Berry phase terms as we will study the critical properties
of $2+1$ D antiferromagnets as approached from the N\`eel phase.

The $O(3)$ nonlinear sigma model action (\ref{action}) has a classical
``ground state'' or lowest energy state with N\`eel order
corresponding to a constant magnetization. The equations of motion
that follow from the action have approximate time dependent solutions,
corresponding to Goldstone spin wave excitations. The equations of
motion, in 2+1 D only, also have exact static solitonic solutions of
finite energy\cite{polya1}
    \beq 
      E = \frac{\rho_s}{2} \int d^2\vec x (\partial_i\vec n)^2 = 4\pi
      \rho_s \;.
    \eneq
These solitons are called skyrmions\cite{skyrme}. 

Skyrmions are of a topological nature as they are characterized by the
integer winding number
    \beq
      \label{q1} 
      q=\frac{1}{8 \pi} \int d^2 x \epsilon^{i j} \vec n \cdot \left(
      \partial_i \vec n \times \partial_j \vec n \right) \;.
    \eneq
These configurations consist in the order parameter rotating an
integer number of times as one moves from infinity toward a fixed but
arbitrary position in the plane. Since two dimensional space can be
thought of as an infinite 2 dimensional sphere where the magnetic
moments live, the excitations fall in homotopy classes of a 2D sphere
into a 2D sphere: $S^2 \rightarrow S^2$\cite{polya1}. Skyrmions rotate
at finite length scales but relax into the N\'eel state far away: $
\lim_{|\vec x| \rightarrow \infty} \vec n=(0,0,-1)$. They have a
directionality given by the direction of the N\`eel order they relax
to at infinity. The skyrmion number is a conserved quantum number as
it is the zeroth component of the current $J^\mu = (1/8\pi)
\epsilon^{\mu\nu\sigma} \vec n \cdot \partial_\nu\vec n \times
\partial_\sigma\vec n$ which can easily be checked to be conserved
$\partial_\mu J^\mu=0$.

In order to study skyrmion properties more conveniently, we use a very
useful way of describing the $O(3)$ nonlinear sigma model, which is
through the stereographic projection\cite{polya1,gross1}:
    \beq
      \label{stereo}
      n^1 + i n^2 = \frac{2w}{|w|^2 + 1} \,,
      n^3=\frac{1-|w|^2}{1+|w|^2} \,, w=\frac{n^1 + i n^2}{1+n^3} \; .
    \eneq
In terms of $w$ the nonlinear $\sigma$-model action is 
    \beq 
      \nonumber S_E [w] = \frac{2\Lambda}{g_\Lambda}\int d\tau d^2x
      \frac{\partial^\mu w \partial_\mu w^*}{ (1 + |w|^2)^2}=
    \eneq
    \beq 
      \label{nlsmlag}
      \frac{2\Lambda}{g_\Lambda}\int d\tau d^2x \frac{\partial_0 w
      \partial_0 w^* - 2 \partial_z w \partial_{z^*} w^* - 2
      \partial_{z^*} w \partial_z w^*}{(1 + |w|^2)^2} \; ,
    \eneq
where $z = x + i y$ and $z^* = x -i y$ is its conjugate, and
$\Lambda/g_\Lambda=\rho_s$. $g_\Lambda$ is the microscopic Goldstone
coupling constant defined at the microscopic cutoff scale
$\Lambda$\cite{halp,polya2}. The classical equations of motion which
follow by stationarity of the classical action are
    \beq
      \nonumber
      \partial_0^2 w - 4\partial_z\partial_{z^*} w = \frac{2 w^*}{1 +
      |w|^2} \left[ (\partial_0 w)^2 - 4 \partial_z w \partial_{z^*} w
      \right] = 0
    \eneq
When the system N\`eel orders, $\vec n$, or equivalently $w$, will
acquire an expectation value: $\langle n^a \rangle = \delta^{3a}$,
$\langle w \rangle = 0$.

As mentioned above, besides Goldstones, there are static skyrmion
configurations\cite{polya1,gross1}: $ w = \prod_{i=1}^q \lambda/ (z -
a_i)$ whose topological invariant (\ref{q1}) in terms of the
stereographic variable, $w$, is
    \beq
      \label{q2} 	
      q = \frac{1}{\pi} \int d^2 x \frac{\partial_z w \partial_{z^*}
      w^* - \partial_{z^*} w \partial_z w^*}{ (1 + |w|^2)^2} \,.
    \eneq
This configuration can easily be checked to have charge $q$ and energy
$4\pi q\Lambda/g_\Lambda$. $\lambda^q$ is the arbitrary size and phase
of the configuration and $a_i$ are the positions of the skyrmions that
constitute the multiskyrmion configuration. Similarly, the
multiantiskyrmion configuration can be shown to be
$w=\prod_{i=1}^q\lambda^*/(z^*-a_i^*)$ with charge $-q$ and energy
$4\pi q\Lambda/g_\Lambda$.

We now investigate whether skyrmions and antiskyrmion configurations
are relevant at the quantum critical point. As mentioned above, their
classical energy is $4\pi\Lambda/g_\Lambda$, which is independent of
the size of the skyrmion $\lambda$. On the other hand, in real
physical systems there are quantum and thermal fluctuations. These
renormalize the effective coupling constant of the nonlinear sigma
model and makes it scale dependent. To one loop order the renormalized
coupling constant is
    \beq
      \label{gren}
      g_\mu = \frac{\mu}{\Lambda} \frac{g_\Lambda}{1 -
      (g_\Lambda/2\pi^2)\left(1 - \mu/\Lambda\right)} \,.
    \eneq
Since the skyrmion has an effective size $\lambda$, spin waves of
wavelength smaller than $\lambda$ renormalize the energy of the
skyrmion via the coupling constant renormalization leading to an
energy and Euclidean action $S_E = \beta E$, which are now scale
dependent through the scale dependence of the coupling constant at
scale $\mu = 1/\lambda$. If the system is at temperature $T =
1/\beta$, this temperature sets the size of the skyrmion to be the
thermal wavelength $\lambda = \beta$. The skyrmion Euclidean action is
then
    \beq
      \label{SE}
      S_E = \frac{8\pi\beta}{\beta g_{1/\beta}} =
      \frac{8\pi\beta\Lambda}{g_\Lambda}\left[ 1 -
      \frac{g_\Lambda}{2\pi^2} \left( 1 - \frac{1}{\beta\Lambda}
      \right) \right]
    \eneq

Having obtained the Euclidean action for skyrmions (\ref{SE}), we now
study its low temperature limit in the N\`eel ordered phase and at the
quantum critical point.  According to the one loop renormalized
coupling constant (\ref{gren}), the quantum critical point occurs when
the renormalized spin stiffness ($\rho_s(\mu)\propto \mu/g_\mu$)
vanishes at long wavelengths ($\mu\rightarrow 0$): $g_\Lambda = g_c =
2\pi^2$. Since the skyrmion gap is $4\pi\mu/g_\mu$, the critical
point corresponds to skyrmion gap collapse. When in the N\`eel ordered
phase, $g_\Lambda < g_c$, the skyrmion Euclidean action (\ref{SE}) is
infinite. Therefore, the probability for skyrmion contributions is
suppressed exponentially at low temperatures, vanishing at zero
temperature. {\it Skyrmions are gapped and hence irrelevant to low
temperature physics in the N\`eel ordered phase}.

At the quantum critical point $g_\Lambda = g_c = 2\pi^2$, the skyrmion
Euclidean action is
    \beq
      S_E = \frac{4}{\pi} \,.
    \eneq
This action is finite and constant at all temperatures and in
particular, it will have a nonzero limit as the temperature goes to 0:
the skyrmion probability is nonzero and constant at arbitrarily low
temperatures and zero temperature. {\it Hence there are skyrmion
excitations at criticality at arbitrarily low energies and
temperatures, including zero at zero temperature. Therefore skyrmion
excitations contribute to quantum critical physics}.

We have seen that skyrmions are relevant at criticality as the
critical point is associated with skyrmion gap collapse and they have
a nonzero probability to be excited at arbitrarily low temperature at
criticality. On the other hand, skyrmions have nonzero conserved
topological number while the ground state has zero skyrmion
number. Absent any external sources that can couple directly to
skyrmion number, they will always be created in equal numbers of
skyrmions and antiskyrmions. Therefore, in order to study the effect
of skyrmions and antiskyrmions we need to include configurations with
equal number of skyrmions and antiskyrmions in the partition function
or path integral. We found a time independent solution to the
equations of motion given by
    \beq
      w_t^{(n)} =
      e^{i\varphi}\tan\left[\left(\frac{\lambda}{z-a}\right)^n +
      \left(\frac{\lambda^*}{z^* - a^*}\right)^n + \frac{\theta}{2}
      \right]
    \eneq

where $\lambda$ is the size of the configuration, $\theta$ and
$\varphi$ are the arbitrary directions of the configurations, $a$ is
the arbitrary position of the configuration and $n$ is an
integer. This configuration is topologically trivial because it has
$q=0$ as obtained from (\ref{q1}). On the other hand, it is composed
of arbitrary superpositions of equal numbers of skyrmions and
antiskyrmions with $q=\pm n$, i.e. the precise superpositions we need
to sum over in the path integral for the $q=0$ sector. Since this
configurations is made of topologically nontrivial skyrmions and
antiskyrmions, we dub it a topolon. While the argument of the tangent
is obviously a sum of an $n$ skyrmion and an $n$ antiskyrmion, it
appears to not be a fully general one as all the skyrmions are at the
same position. By starting with a fully general skyrmion configuration
and making a change of variables to an effective ``center of mass''
coordinate, it follows that the results are the same as having all
skyrmions at the same place.

The topolon with spatial and temporal size $\lambda$ has Euclidean
action (\ref{nlsmlag})
    \beq
      S_E^t = \int_0^\lambda d\tau\frac{8\pi}{\lambda g_{1/\lambda}}
      \left( \lambda\Lambda \right)^{2n} \simeq
      \frac{8\pi\lambda\Lambda}{g_\Lambda} \left( \lambda\Lambda
      \right)^{2n} + \mathcal O(g_\Lambda^0)
    \eneq
The partition function including topolon configurations is given by
$\mathcal Z = \sum_{n=0}^\infty \mathcal Z_n$ where

    \beq
      \mathcal Z_0 = \int \frac{\mathcal D \nu \mathcal D\nu^*}{(1 +
      |\nu|^2)^2} e^{-S_E[\nu]}
    \eneq
is the usual partition function for the nonlinear sigma model with no
topolons and only the spin wave like fields $\nu$ and $S_E$ is the
Euclidean action for the nonlinear sigma model in terms of the
stereographic projection variables (\ref{nlsmlag}). We also have that
    \beq
      \nonumber
      \mathcal Z_{n\neq 0} = \int \frac{\mathcal D\nu \mathcal D
      \nu^*}{(1 + | w_t^{(n)} + \nu|^2)^2} \frac{d^2a}{A}
      \frac{d\Omega}{4\pi} \frac{d\lambda}{1/\Lambda} \;
      e^{-S_E[w_t^{(n)} + \nu]} \,.
    \eneq
The $Z_{n\neq 0}$ is the path integral with the $n$ topolons with spin
waves $\nu$. Besides integrating over the spin wave configurations, we
must integrate over the topolon parameters: its size $\lambda$
normalized to the lattice spacing $1/\Lambda$, its position $a$
normalized to the area $A$ of the system, and over the solid angle of
its orientation normalized to $4\pi$.

In order to renormalize the nonlinear sigma model, we integrate the
spin wave degrees of freedom $\nu$ with momenta between the
microscopic cutoff $\Lambda$ and a lower cutoff or renormalization
scale $\mu$\cite{polya2}. We also integrate topolons of size between
the microscopic minimum length $1/\Lambda$ and a new larger
renormalization length $1/\mu$. We thus obtain a double expansion in
$(1 - \mu/\Lambda)$ and the coupling constant which leads to the
renormalized action, the renormalized spin stiffness and the beta
function
    \begin{align}
      \begin{aligned}
	\nonumber
	&S_{\text{ren}} = \frac{2\mu}{g_\mu} \int d^3x
	\frac{\partial_\mu\nu \partial_\mu\nu^*}{(1 + |\nu|^2)^2}
	\simeq \frac{2\Lambda}{g_\Lambda} \int d^3x \partial_\mu\nu
	\partial_\mu\nu^* \times \\
	&\left\{ 1 - \left(1 - \frac{\mu}{\Lambda}\right)
	\frac{g_\Lambda}{2\pi^2} + \left( 1 - \frac{\mu}{\Lambda} \right)
	\frac{1}{3(e^{8\pi/g_\Lambda} - 1)} \right\} \\ 
	&\rho_s(\mu) = \frac{\mu}{g_\mu} = \frac{\Lambda}{g_\Lambda}
	\times \\
	&\left\{ 1 - \left(1 - \frac{\mu}{\Lambda}\right)
	\frac{g_\Lambda}{2\pi^2} + \left( 1 - \frac{\mu}{\Lambda} \right)
	\frac{1}{3(e^{8\pi/g_\Lambda} - 1)} \right\} \\
	&\beta (g) = \mu \frac{\partial g}{\partial\mu} \Big|_{\Lambda
	= \mu} = g - \frac{g^2}{2\pi^2} + \frac{g}{3(e^{8\pi/g} - 1)} \,.
      \end{aligned}
    \end{align}
The last term in the spin stiffness and in the beta function is the
contribution from the topolons and the rest is the contribution of the
spin waves. The coupling constant at the quantum critical point $g_c
\simeq 23.0764$ is obtained from $\beta(g_c) = 0$.

The correlation length at scale $\mu$ is given by\cite{polya2}
    \beq
      \xi \sim \mu^{-1} \exp \left[ \int_{g_c}^{g_\mu} \frac{dg}{\beta
      (g)} \right] \sim \mu^{-1} \left( g_c - g_\mu \right)^{1/\beta
      '(g_c)} \,.
    \eneq
The correlation length exponent is $\nu = -1/\beta ' (g_c) \equiv
-(d\beta/dg|_{g=g_c})^{-1}$. The correlation length exponent with
topolon contributions evaluates to $\nu=0.9297$. The $d=2+\epsilon$
expansion of the $O(N)$ vector model, which agrees with the $1/N$
expansion for large $N$, gives $\nu=0.5$\cite{zj}. We note that our
value is larger than the accepted numerical evaluations of critical
exponents in the Heisenberg model, $\nu=0.71125$\cite{nu}, but about
as close to this accepted Heisenberg value than the $2+\epsilon$
expansion or the $1/N$ expansion. We conjecture that the difference
between our value and the Heisenberg value is real and attributable to
quantum critical degrees of freedom.

Goldstone renormalizations of the ordering direction $\sigma = n_3$,
and hence of the anomalous dimension $\eta$, are notoriously
inaccurate. The one loop approximation leads to a value of $\eta=2$,
thousands of percent different from the accepted numerical value of
$\eta\simeq 0.0375$\cite{nu}. The large $N$ approximation, which sums
bubble diagrams, is a lot more accurate. To order $1/N$ one obtains
$\eta = 8/(3\pi^2 N) \simeq 0.09$ for $N=3$. We now calculate the
value of $\eta$ from topological nontrivial configurations
    \begin{align}
      \begin{aligned}
	\nonumber
	&\langle n_3^2 \rangle = Z = 1 - \langle n_1^2 + n_2^2 \rangle
	= 1 - \left\langle \frac{4 |w|^2}{(1 + |w|^2)^2} \right\rangle
	\\ 
	&\simeq 1 - \frac{2}{3} \frac{1 - \mu/ \Lambda}{
	e^{8\pi/g_\Lambda} - 1 } \Rightarrow \eta (g) =
	\frac{\mu}{Z}\frac{\partial Z}{\partial\mu}\Big |_{\Lambda =
	\mu} \simeq \frac{2}{3 (e^{8\pi/g} - 1)}
      \end{aligned}
    \end{align}
For the anomalous dimension at the quantum critical point we obtain
$\eta(g_c) = 0.3381$. On the other hand, we have seen that spin wave
contributions tend to give quite large and nonsensical values of
$\eta$. In fact so large as to wash out the momentum dependence of the
propagator. Hence, to calculate $\eta$, spin wave contributions prove
to be tough to control. Our calculation gives a value quite larger
than the accepted numerical value.  We have recently calculated
\cite{us} the unique value of $\eta$ that follows from quantum
critical fractionalization into spinons and find $\eta = 1$. While our
value obtained from topolons is far from 1, it is a lot closer than
the accepted numerical Heisenberg value and the $1/N$ value.

\end{document}